\documentclass[conference]{IEEEtran}
\IEEEoverridecommandlockouts
\pdfoutput=1
\usepackage{cite}
\usepackage{amsmath,amssymb,amsfonts}
\usepackage{algorithmic}
\usepackage{graphicx}
\usepackage{textcomp}
\usepackage{xcolor}

\usepackage{orcidlink}
\usepackage[absolute,overlay]{textpos}

\def\BibTeX{{\rm B\kern-.05em{\sc i\kern-.025em b}\kern-.08em
    T\kern-.1667em\lower.7ex\hbox{E}\kern-.125emX}}

\begin{document}

\begin{textblock*}{10cm}(16.5cm,0cm)
\includegraphics[width=3cm]{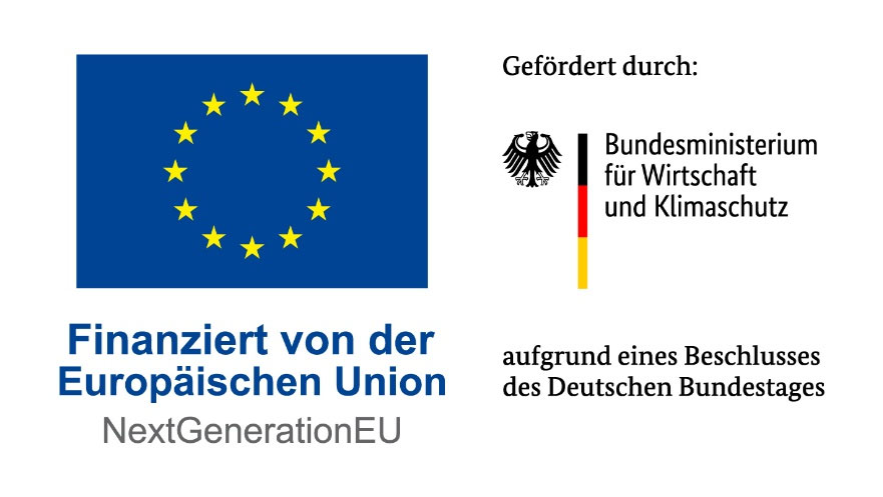}
\end{textblock*}

\title{GMB-ECC: Guided Measuring and Benchmarking of the Edge Cloud Continuum\\
\thanks{Research is funded by T-Systems International GmbH.}
}

\author{\IEEEauthorblockN{1\textsuperscript{st}  Brian-Frederik Jahnke\orcidlink{0000-0002-7995-6472}}
\IEEEauthorblockA{\textit{Fraunhofer ISST} \\
Dortmund, Germany \\
brian-frederik.jahnke@isst.fraunhofer.de
}
\and
\IEEEauthorblockN{2\textsuperscript{nd} Rebecca Schmook\orcidlink{0009-0008-8612-1394}}
\IEEEauthorblockA{\textit{Fraunhofer ISST} \\
Dortmund, Germany \\
rebecca.schmook@isst.fraunhofer.de
}
\and
\IEEEauthorblockN{3\textsuperscript{rd} Falk Howar\orcidlink{0000-0002-9524-4459}}
\IEEEauthorblockA{\textit{TU Dortmund} \\
Dortmund, Germany \\
falk.howar@tu-dortmund.de
}}

\maketitle

\begingroup
\let\clearpage\relax
\clearpage{}%
\section{abstract}
In the evolving landscape of cloud computing, optimizing energy efficiency across the edge-cloud continuum is crucial for sustainability and cost-effectiveness. We introduce GMB-ECC, a framework for measuring and benchmarking energy consumption across the software and hardware layers of the edge-cloud continuum. GMB-ECC enables energy assessments in diverse environments and introduces a precision parameter to adjust measurement complexity, accommodating system heterogeneity. We demonstrate GMB-ECC's applicability in an autonomous intra-logistic use case, highlighting its adaptability and capability in optimizing energy efficiency without compromising performance. Thus, this framework not only assists in accurate energy assessments but also guides strategic optimizations, cultivating sustainable and cost-effective operations.
\clearpage{}%

\begin{IEEEkeywords}
Edge–Cloud Continuum, Energy Consumption Measurement, Energy Efficiency Benchmarking, Holistic Optimization, Sustainability
\end{IEEEkeywords}

\clearpage{}%
\section{Introduction}
\textbf{Motivation.} The increasing demand for cloud services requires greater computing power, leading to higher energy consumption across the edge-cloud continuum, including data centers, edge devices, and network components \cite{10.1145/2742488,   bundestag2022}. Major cloud providers such as Amazon Web Services, Microsoft Azure, and Telekom AG strive to deliver high-quality services while managing operational costs. Achieving energy efficiency is crucial for sustainability and cost-effectiveness, as optimized resource utilization leads to significant energy savings, reduced operational costs, and improved system performance \cite{BELOGLAZOV201147}. Utilizing data analytics and real-time monitoring improves our understanding of resource usage patterns, enabling dynamic adjustments that further optimize energy consumption \cite{GAI2018126}.

\textbf{Problem Statement.} However, optimizing energy efficiency across the expansive and diverse edge-cloud continuum is a challenge. Current manual optimization processes are often labor-intensive, error-prone, and struggle to effectively manage the dynamic and heterogeneous nature of these environments. Local optimizations can sometimes degrade overall system performance rather than enhance it, due to the complex interactions between components.
There is a pressing need for a framework that guides where to focus optimization efforts and quantifies the impact of these optimizations on the entire system. Such a framework should integrate hardware and software practices, offering scalable and sustainable solutions to improve energy efficiency across all layers of the edge-cloud continuum. This approach is required for achieving significant energy savings without compromising system performance.

\begin{figure}
    \centering
    \includegraphics[width=1.0\linewidth]{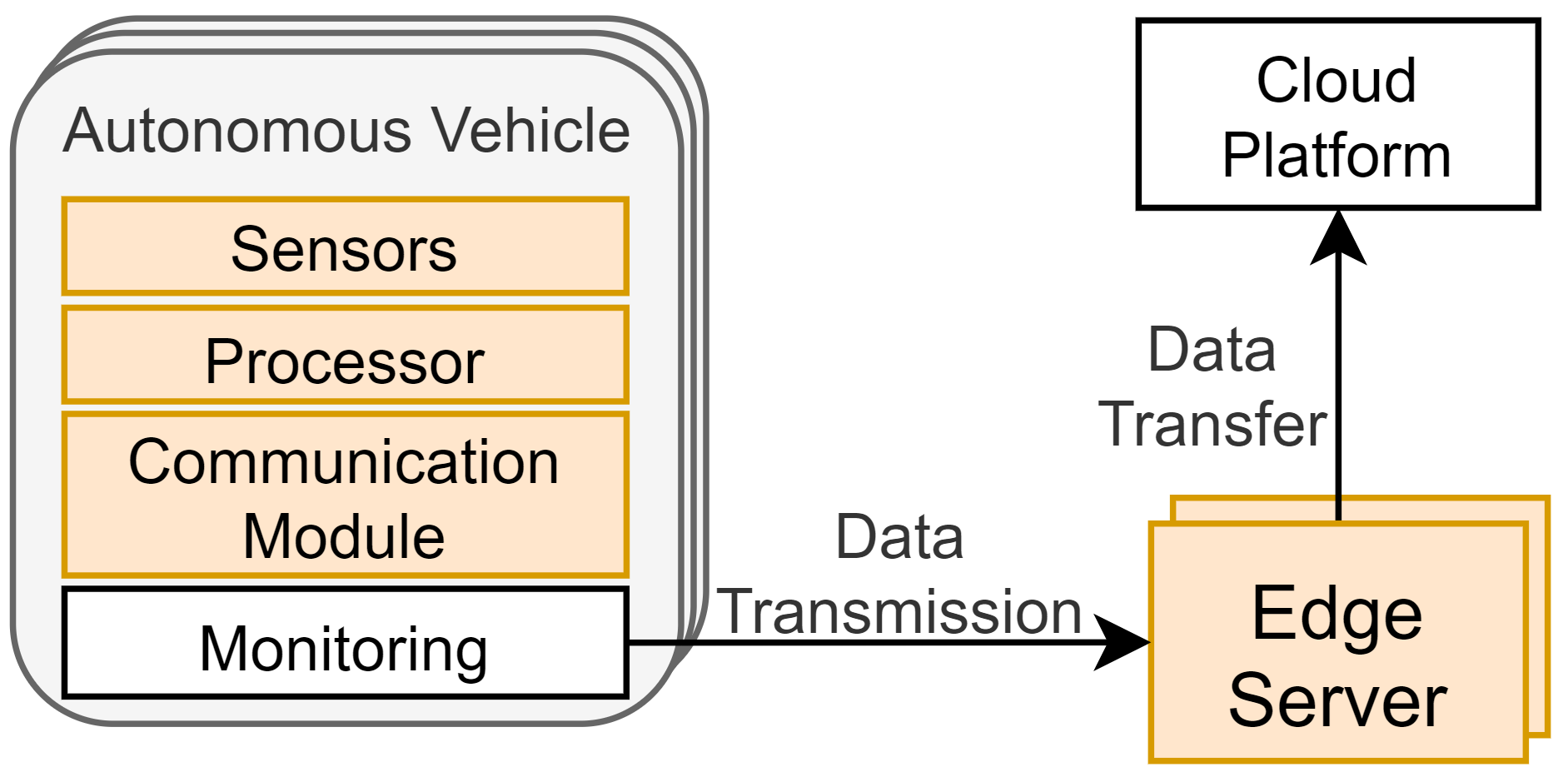}
    \caption{\textbf{Energy consumption data flow.} The energy is measured on the autonomous vehicles and sent to edge computing via the network. The edge computing and network energy is derived from utilization and measurement. All data is sent to the cloud for further processing and analysis.}
    \label{fig:overview-use-case}
\end{figure}

\textbf{State-of-the-Art Approaches.}
Most energy measurement methods focus on cloud data centers, not observing the spectrum that includes IoT devices, edge computing, and networks. Therefore, these techniques may not effectively identify bottlenecks or the largest energy consuming components. Current methodologies target isolated measures of energy efficiency, such as energy-aware offloading, resource allocation and workload management for performance enhancement. For instance, while Cong et al. examine a mobile edge cloud context, their emphasis is on computation offloading methods \cite{Cong.2020}. A survey on energy-aware edge computing revealed that, different from simple cloud settings, energy efficiency in cloud edge is largely left uninvestigated due to the more complex relations in those settings \cite{Jiang.2020}. Although methods for energy-efficient services and applications were reviewed, the survey does not include energy consumption measurements or benchmarks. Traditional solutions are statically designed, lacking the ability to dynamically adapt to varying workloads and falling short in scalability \cite{Ahvar.2021}. This leads to maintenance challenges that are hard to manage at scale \cite{Aouedi.2024}.

Furthermore, they lack integration with existing orchestration technologies and fail to address the edge-cloud continuum. Recent efforts strive for accurate measurement techniques and frequent data collection at shorter intervals, progressing towards real-time energy consumption data \cite{Ismail.2024}. However, these approaches do not establish key performance indicators for energy efficiency or standardized benchmarking methods, rendering them incompatible with energy measurement across the continuum.

\textbf{Challenges.} Developing a framework for measuring and benchmarking energy consumption across the edge-cloud continuum presents several challenges:

\begin{itemize}
    \item \textit{Heterogeneity of Components:} The edge-cloud continuum comprises a wide variety of hardware and software components with different characteristics and energy profiles. Creating a unified model that accurately represents this heterogeneity is complex.
    \item \textit{Scalability of Measurement Techniques:} Implementing fine-grained energy measurements can introduce significant overhead, especially in large-scale systems. The framework must balance measurement precision with scalability.
    \item \textit{Integration with Existing Systems:} Seamless integration with current orchestration and management tools is needed without requiring extensive modifications to existing infrastructures.
    \item \textit{Dynamic Workloads and Environments:} The energy consumption patterns in cloud environments are highly dynamic due to fluctuating workloads. The framework must account for these variations to provide accurate measurements and meaningful benchmarks.
\end{itemize}

\textbf{Our Proposed Solution.}
We introduce the Guided Measuring and Benchmarking framework for Energy Consumption across the Edge-Cloud Continuum (GMB-ECC), a solution designed to systematically assess and optimize energy usage within heterogeneous edge-cloud environments. By leveraging existing hardware and software practices, GMB-ECC facilitates energy assessments without the need to develop new tools or algorithms. By integrating these practices, we provide a holistic optimization method across all technology layers, accommodating the diverse environments of edge-cloud systems.

A key feature of GMB-ECC is the precision parameter, which allows for adjustment of measurement granularity to balance detail and computational overhead. This enables an accurate reflection of the diverse characteristics of components within the edge-cloud continuum, promoting efficient energy monitoring and management.

We implemented GMB-ECC in an autonomous intra-logistic application, resulting in notable energy efficiency gains while maintaining system performance, demonstrating the applicability of GMB-ECC. By unifying existing energy efficiency practices, GMB-ECC enables providers to accurately assess energy consumption, identify areas for improvement, and implement optimizations leading to more cost-effective operations.

\textbf{Contributions.}
Our contributions include the following:

\begin{itemize}
    \item The development of the GMB-ECC framework, which provides an approach to measuring and benchmarking energy consumption across the edge-cloud continuum, integrating both hardware and software practices while being adaptable to various deployment scenarios.
    \item The introduction of an energy efficiency analysis methodology that encompasses state representation, efficiency gap calculation, and prioritization of components for optimization, aiding in the identification and resolution of inefficiencies within the system.
    \item The establishment of standardized benchmarking metrics and categorization criteria for evaluating energy efficiency, facilitating consistent assessment and comparison across diverse systems and components.
    \item An implementation and demonstration of the GMB-ECC framework through a current application example, showcasing its adaptability and effectiveness in uncovering optimization opportunities.
    \item The enhancement of the framework with a precision parameter, which allows for adjustable measurement granularity, ensuring that energy efficiency metrics accurately represent system heterogeneity while maintaining construct validity.
\end{itemize}

\textbf{Outline.}
In the following we present our framework in Section \ref{sec:methodology} and demonstrate our framework on a running application example in Section \ref{sec:evaluation}. We review related work in Section \ref{sec:related_work} and discuss our framework's limitations and future work in Section \ref{sec:limitations_and_future_work} and lesson learned in Section \ref{sec:lessons_learned}. We conclude the paper in Section \ref{sec:conclusion}.
\clearpage{}%
\clearpage{}%
\section{Methodology}\label{sec:methodology}
In this section, we present our proposed methodology for measuring and benchmarking energy consumption across the edge-cloud continuum. The methodology is divided into three sequential steps: State Representation, Energy Efficiency Analysis, and Benchmarking.

\subsection{State Representation}
We represent the edge-cloud continuum (e.g. Figure \ref{fig:component-model}) as a weighted directed acyclic graph \( G_s = (K_s, E_s, w_s) \) (Figure \ref{fig:service-energy-measurement-model}), where:

\begin{itemize}
    \item $s$: The state of the system where $S$ is the set of all possible system states.
    \item \( K_s \) is a set of nodes, each node representing a component in the edge-cloud continuum in state $s$.
    \item \( E_s \) is a set of directed edges, represented as ordered pairs \( (u, v) \) where \( u, v \in K_s \) and $u \neq v$. An edge indicates that one component is utilizing another.
    \item \( w_s: E_s \rightarrow [0, 1] \) is a weight function such that each weight \( w_s(u, v) \) indicates the degree of utilization of one component by another.
\end{itemize}

\begin{figure}
    \centering
    \includegraphics[width=1.0\linewidth]{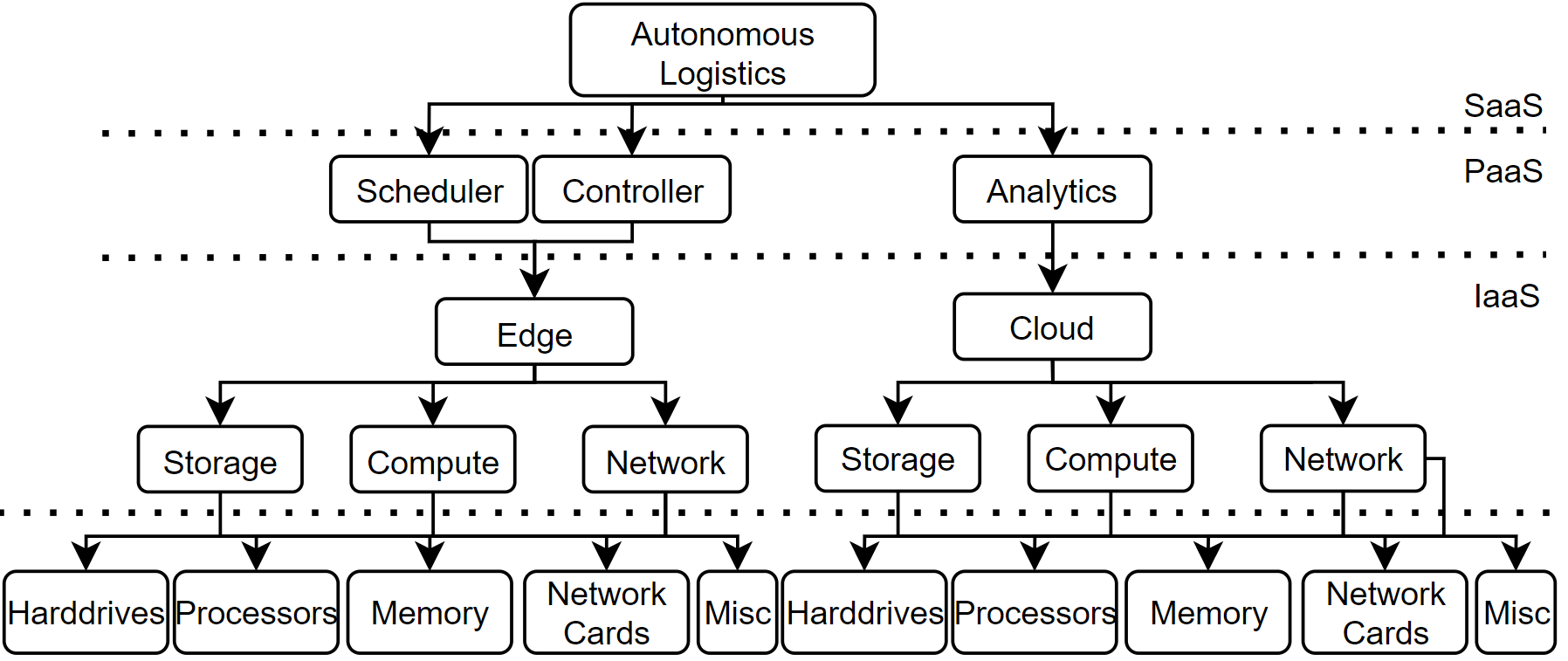}
    \caption{\textbf{Software-Hardware Stack.} Mapping of software processes to underlying hardware resources in the edge-cloud continuum.}
    \label{fig:component-model}
\end{figure}

Each node \( k \in K_s \) is characterized by an efficiency function \( \eta_{k,s} \), for instance the energy efficiency curve, and an associated variance \( \sigma_{k,s}^2 \), which is our measurement precision parameter:

\[
\eta_{k,s}: \text{Utilization} \rightarrow \text{Efficiency}, \quad \text{Variance: } \sigma_{k,s}^2
\]

with \(\eta: [0,1] \rightarrow [0,1].\)

The edge function \( f_s \) is defined as:

\[
f_s(u, v) =
\begin{cases} 
w_s(u, v), & \text{if } (u, v) \in E_s\\
0, & \text{if } (u, v) \notin E_s
\end{cases}
\]

For all components \( u \in K \), the weights of the incoming and outgoing edges are, respectively, restricted by :

\[
\sum_{v \in K_s} f_s(u, v) = 1;
\]

We further define:

\textbf{Measurable Components:} Measurable components \( L_s \subseteq K_s \) are the set of components with no outgoing edges:

\[
L_s = \{ k \in K_s \mid \nexists (k, u) \in E_s \}
\]

These components $l\in L_s$ include for example sensors, processors, network devices, and storage units, whose energy efficiency function \( \eta_{l,s} \) and variance \( \sigma_{l,s}^2 \) are known.

\textbf{Composite Components:} Aggregates composed of multiple measurable components which are the set of components \( I_s \subseteq K_s \setminus L_s \) with outgoing edges:

\[
I_s = \{ k \in K_s \mid \exists (k, u) \in E_s \}
\]

For a composite component \( n\in I_s \), its function \( \eta_{n,s} \) is constructed by aggregating the functions of its connected nodes $K_{n,s}$ (Figure \ref{fig:service-energy-measurement-model}), considering an error term \(\epsilon_n\) to account for unconsidered components (Figure \ref{fig:precision}):

\begin{align}\label{eq:composite_efficiency}
\eta_{n,s} = \sum_{c \in K_{n,s}} f_s(c, n) \cdot \eta_{c,s} + \epsilon_{n,s}
\end{align}

The variance \( \sigma_{n,s}^2 \) for component \( n \) in state $s$ is constructed as:

\begin{align}\label{eq:composite_variance}
\sigma_{n,s}^2 = & \sum_{c \in K_{n,s}} f_s(c, n)^2 \cdot \sigma_{c,s}^2 \nonumber \\
             & + 2 \sum_{\substack{c < d \\ c, d \in K_{n,s}}} f_s(c, n) \cdot f_s(d, n) \cdot \sigma_{c,d, s}
\end{align}
\begin{figure}[t]
    \centering
    \includegraphics[width=1\linewidth]{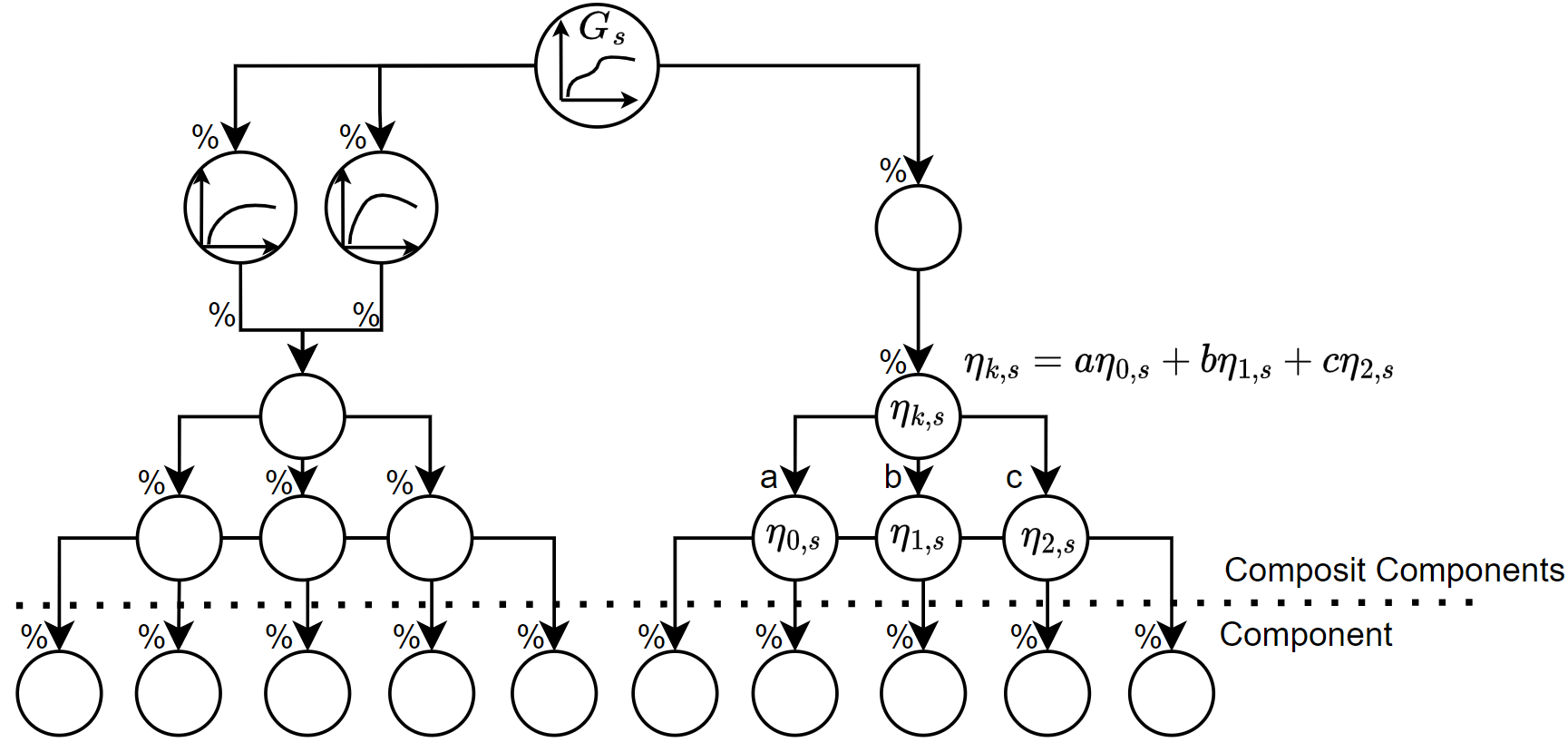}
    \caption{\textbf{Energy efficiency graph.} Showing components and their interrelations within the system, resulting in the composite energy efficiency curves for the whole edge-cloud continuum example sketched in Figure~\ref{fig:service-energy-measurement-model}.}
    \label{fig:service-energy-measurement-model}
\end{figure}
where \( f_s(c, n) \) is the weight indicating the utilization proportion of component \( c \) within \( n \), and \(\sigma_{c,d,s}\) is the covariance between the efficiencies of components \( c \) and \( d \) in state $s$.

These sets are distinct, meaning:

\[
L_s \cap I_s = \emptyset
\]

Therefore, all components of \( L_s\) need to be known to construct all unknown components in the edge cloud continuum.

\subsection{Energy Efficiency Analysis with Error Consideration}
The energy efficiency of each component is analyzed by mapping current utilization against theoretical optimal efficiency, as illustrated in Figure \ref{fig:efficiency_curve_to_benchmark}. This process identifies efficiency gaps critical for optimization within the edge-cloud continuum.

\subsubsection{Merging Multiple Energy Graphs}
To create a unified representation $G_{S^\ast}$ of multiple graphs defined over all observed states $s\in S^\ast \subseteq S $, for example at different timestamps, we perform the following steps:

\begin{itemize}
    \item \textbf{Unified Node Set:} Define \( K_{S^\ast} \) as the union of all nodes across states $s$:
    \[
    K_{S^\ast} = \bigcup_{s\in S^\ast} K_{s}
    \]

    \item \textbf{Unified Edge Set:} Define \( E_{S^\ast} \) as the union of all edges:
    \[
    E_{S^\ast} = \bigcup_{s\in S^*} E_{s}
    \]

    \item \textbf{Weight Aggregation:} For each edge \( (u, v) \in E_{S^\ast} \), calculate the aggregated weight \( w_{S^\ast}(u, v) \):
    \[
    w_{S^\ast}(u, v) = \frac{1}{|S^\ast|} \sum_{s \in S^\ast, u, v \in K} f_{s}(u, v)
    \]

    \item \textbf{Constructing the Merged Graph:} The merged graph \( G_{S^\ast} = (K_{S^\ast}, E_{S^\ast}, w_{S^\ast}) \) is formed using the aggregated nodes, edges, and weights over all states $S^\ast$. The composite efficiency curve \ref{eq:composite_efficiency} and variance \ref{eq:composite_variance} are defined with $s \in S^\ast$ for components $k$ with:
    \[
    \eta_{k, S^\ast} = \frac{1}{|S^\ast|} \sum_{s \in S^\ast} \eta_{k, s}
    \] 
    and 
    \[
    \sigma_{k, S^\ast}^2 = \frac{1}{|S^\ast|} \sum_{s \in S^\ast} \sigma_{k, s}^2
    \] 
    
\end{itemize}
\begin{figure}[t]
    \centering
    \includegraphics[width=1\linewidth]{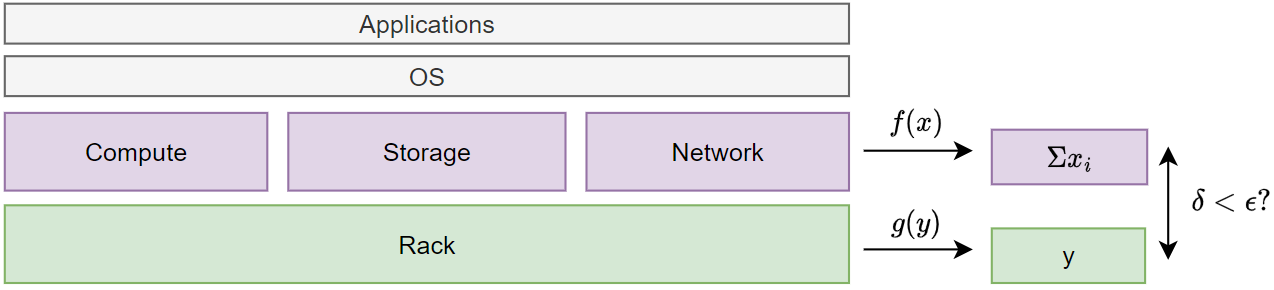}
    \caption{\textbf{Measurement and validation process.} Measurements at different layers are compared and validated to ensure accuracy.}
    \label{fig:precision}
\end{figure}
\subsubsection{Determining Theoretical Maximum Efficiency}
For each component \( k \in K_{S^\ast} \), the utilization level \( u_{opt, k, S^\ast} \) that achieves maximum energy efficiency is determined by maximizing its function:

\begin{equation}\label{eq:e_max_k}
u_{opt, k, S^\ast} = \arg\max_{u \in [0,1]} \eta_{k,S^\ast}(u)
\end{equation}

\subsubsection{Calculating the Efficiency Gap}
The efficiency gap \(\Delta \eta_{max, k, S^\ast}\) is defined as the difference between the maximum efficiency and the current efficiency:

\begin{align}\label{eq:delta_e_max_k}
\Delta \eta_{max, k, S^\ast} = & \min(1, \nonumber \\
                       & | \eta_{k, S^\ast}(u_{opt, k, S^\ast}) - \eta_{k, S^\ast}(u_{current, k, S^\ast}) | \\
                       & + \sqrt{\sigma_{opt, k, S^\ast}^2 + \sigma_{current, k, S^\ast}^2}) \nonumber 
\end{align}

where \( u_{current, k, S^\ast} \) is the current utilization level and \(\sigma_{opt, k, S^\ast}^2\) and \(\sigma_{current, k, S^\ast}^2\) are variances at optimal and current utilization, respectively. The efficiency gap is capped at 1, describing the maximal possible efficiency gap.

\subsection{Model Fitness}
To ensure the fitness of the model to describe the edge cloud continuum, we derived, from Rosenblatt's learning rule \cite{rosenblatt}, the following protocol to adjust the model:

\begin{enumerate}
    \item If the variance of the target node for benchmarking is acceptable, then do nothing.
    \item If the variance of the target node for benchmarking is smaller than required, find nodes in the subgraphs to prune or simplify, preferring nodes with high measurement cost and low variance, reducing measurement costs.
    \item If the variance of target node for benchmarking is too high, find nodes in the subgraphs with high variance, preferring nodes with low measurement cost and high variance, reducing model variance.
\end{enumerate}

\subsection{Benchmarking}
Benchmarking categorizes components based on their aggregated efficiency gaps to prioritize optimization efforts effectively. We employ thresholds \( a \) and \( b \) to classify components into three categories:

\begin{enumerate}
    \item \textbf{Well-Tuned System (\(\Delta\eta_{max, k, S^\ast} < a\))}: Components operate near optimal efficiency.
    \item \textbf{Partially Optimized System (\(a \leq \Delta\eta_{max, k, S^\ast} \leq b\))}: Components offer opportunities for optimization.
    \item \textbf{Misconfigured System (\(\Delta \eta_{max, k, S^\ast} > b\))}: Components require immediate optimization.
\end{enumerate}

\clearpage{}%
\clearpage{}%
\section{Evaluation}\label{sec:evaluation}
\begin{figure}[t]
    \centering
    \includegraphics[width=1\linewidth]{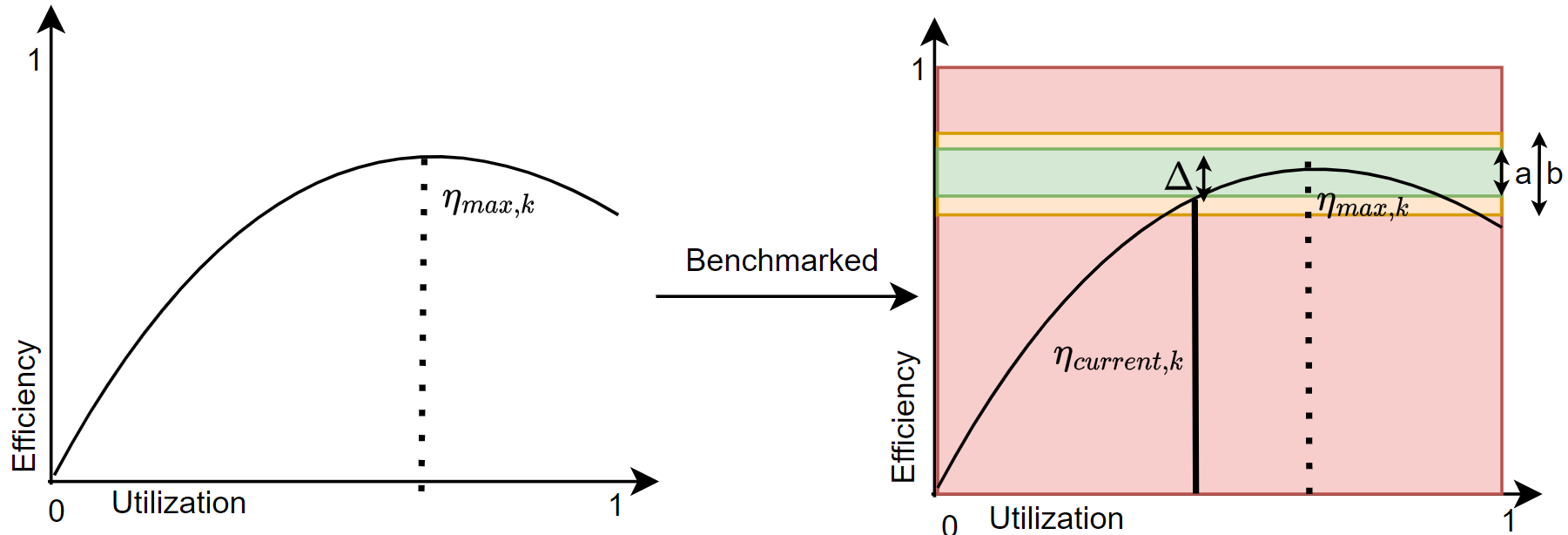}
    \caption{\textbf{Efficiency curve for benchmarking.} Current utilization is located on the efficiency curve and related to the theoretical optimal efficiency, used for benchmarking the component.}
    \label{fig:efficiency_curve_to_benchmark}
\end{figure}

To demonstrate the practical application and effectiveness of the GMB-ECC framework, we applied it to an autonomous intra-logistics use case based on synthetic data. The GMB-ECC framework allowed us to identify whether local changes would have positive or negative impacts on a global scale within our use case. Furthermore, it enabled us to determine if the effort required for local optimizations was justified by the resultant impact on the overall system.

It is important to note that all data presented is synthetic, generated to illustrate concepts and support understanding of how the GMB-ECC workflow operates. This data is intended for following along in the application of the framework.

\subsection{Use Case: Autonomous Intra-Logistics}

We consider a logistics company operating a fleet of autonomous vehicles (AVs) for goods delivery (Figure \ref{fig:component-hierarchy}). Each AV is equipped with on-board sensors, including LiDAR sensors (e.g., Velodyne VLP-16~\cite{velodyne_lidar}), cameras (e.g., FLIR Blackfly S~\cite{flir_camera}), and GPS modules (e.g., u-blox NEO-M8U~\cite{ublox_gps}), for environmental perception and navigation. An edge computing device, e.g. the NVIDIA Jetson AGX Xavier module~\cite{nvidia_xavier}, handles real-time data processing and decision-making. Communication modules, such as 5G modems like the Quectel RM500Q-GL~\cite{quectel_5g}, enable high-speed data transfer to edge servers and cloud platforms.

The edge-cloud continuum encompasses the AVs, edge servers at distribution centers, and cloud-based analytics platforms hosted on infrastructure like Dell PowerEdge R740 servers~\cite{dell_poweredge}. The company's objective is to optimize energy consumption across this continuum to reduce operational costs and enhance sustainability.

Figure~\ref{fig:overview-use-case} illustrates the energy consumption data flow within the system. It highlights the critical points where energy is measured and how data traverses from AVs to the cloud, emphasizing areas for potential optimization.

\subsection{Application of the GMB-ECC Framework}

We applied the GMB-ECC framework iteratively, systematically identifying critical components, determining appropriate measurement methods, and implementing optimizations.

\subsubsection{Defining Energy Efficiency Goals}

Our primary goals were to:

\begin{itemize}
    \item \textbf{Minimize Energy Consumption}: Reduce overall energy usage of AVs and edge devices to improve operational efficiency.
    \item \textbf{Optimize Resource Allocation}: Balance computational tasks between edge and cloud to enhance energy efficiency and performance.
    \item \textbf{Maintain System Responsiveness}: Ensure low latency for real-time decision-making.
\end{itemize}

\subsubsection{State Representation}

We categorized system components into measurable and composite types. Measurable components include sensors, processors, and communication modules. Composite components aggregate these measurable components, forming the AV system, edge servers, and cloud platforms.

\begin{figure}[t]
    \centering
    \includegraphics[width=\linewidth]{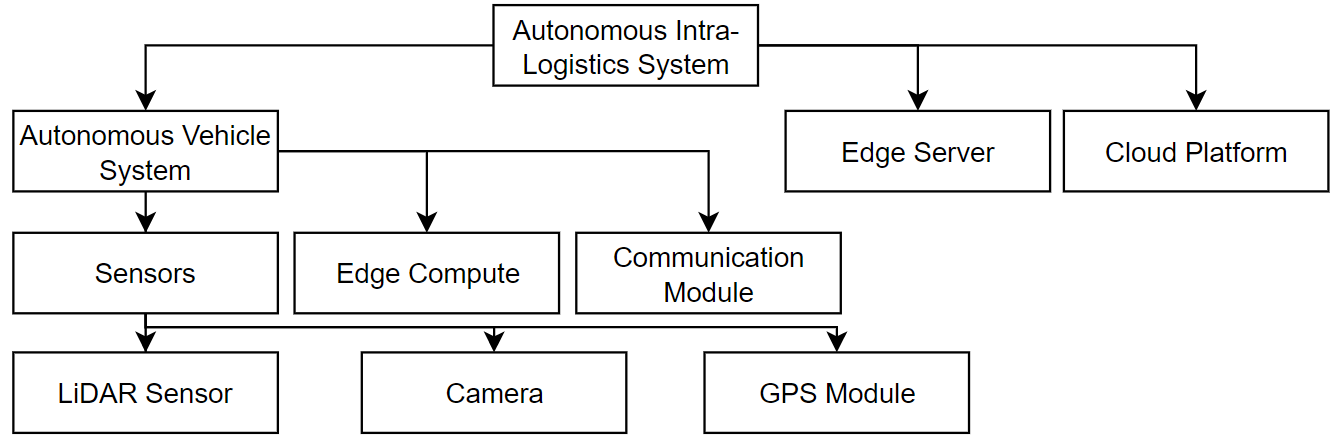}
    \caption{\textbf{System component hierarchy.} Hierarchical representation of system components in the energy efficiency model, showcasing the relationships among AVs, edge servers, and cloud platforms.}
    \label{fig:component-hierarchy}
\end{figure}

Figure~\ref{fig:component-hierarchy} presents the hierarchical structure of the system components. It highlights the interaction between components, essential for constructing the energy efficiency model and identifying key areas for optimization.

\subsubsection{Energy Efficiency Analysis}
The framework allowed us to calculate the efficiency gap ($\Delta \eta_{\text{max},k, S^\ast}$) for each component by comparing current performance to theoretical maximum efficiency. Components with significant efficiency gaps, such as the processor and communication module, were identified as high-priority targets for optimization.

We collected utilization data over a seven-day period using tools like NVIDIA's \texttt{tegrastats}~\cite{tegrastats}, capturing metrics such as CPU/GPU load and network throughput. This data revealed component interactions; for example, Figure~\ref{fig:processor-efficiency} shows the different efficiency curves, highlighting that efficiency peaks at different utilizations, for example the AVs peak efficiency at around 81\%.

Using the methodology outlined earlier, we conducted an energy efficiency analysis focusing on data collection, efficiency modeling, and identifying efficiency gaps.

\begin{figure}[t]
    \centering
    \includegraphics[width=\linewidth]{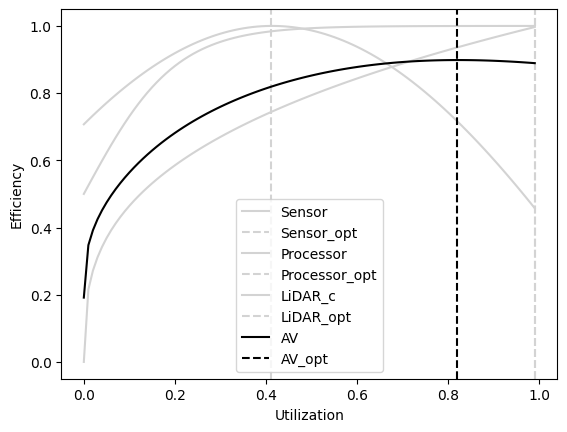}
    \caption{\textbf{Synthetic component efficiency curves.} The curve illustrates energy efficiency and the peak utilization at 81\% of an AV, when constructed as composite component based of three measurable components.}
    \label{fig:processor-efficiency}
\end{figure}

\paragraph{Constructing the Composite Efficiency Model}

Using collected data, we derived connection weights $w_{ij}$ representing the influence of component $i$ on component $j$. This allowed us to build a composite energy efficiency model, capturing interdependencies among components.

\paragraph{Calculating the Efficiency Gap}

We calculated the efficiency gap $\Delta \eta_{k,S^\ast}$ for each component. Components with significant gaps, such as the processor and communication module, were identified for optimization.

\paragraph{Assessing the Worthiness of Optimization Efforts}
We assessed if the effort needed for local changes was warranted by the overall global impact on the system (Figure \ref{fig:energy-consumption}). Consequently, we proceeded.

\subsubsection{Optimization Strategies}

Guided by the GMB-ECC framework, targeted optimization techniques were implemented to reduce $\Delta \eta_{k, S^\ast}$:

\paragraph{Dynamic Voltage and Frequency Scaling (DVFS)}

Applying DVFS to the processor, we adjusted its frequency and voltage based on workload demands~\cite{dvfs_technique}. As shown in Figure~\ref{fig:dvfs-impact}, DVFS reduced the processor's average energy consumption from 25 W to 20 W without compromising performance.

\paragraph{Data Transmission Optimization}

For the communication module, we optimized energy consumption by adjusting data transmission intervals and employing data compression \cite{data_compression_lz4}. Implementing an adaptive transmission algorithm and applying lossless compression reduced the module's energy consumption from 4 W to 3 W.

\subsubsection{Iterative Optimization and Results}

Through iterative optimizations, we refined the system's energy efficiency. Figure~\ref{fig:energy-consumption} illustrates the reduction in energy consumption over two optimization iterations, highlighting significant improvements in the processor and communication module.

\subsection{Results}
The application of the GMB-ECC framework guided us through a systematic and incremental optimization process, ensuring each step was validated and justified in contributing to global system improvements. By leveraging the framework, we effectively identified and prioritized components with significant efficiency gaps, leading to targeted optimizations that enhanced overall system performance. Key improvements achieved include:

\begin{figure}[t]
    \centering
    \includegraphics[width=\linewidth]{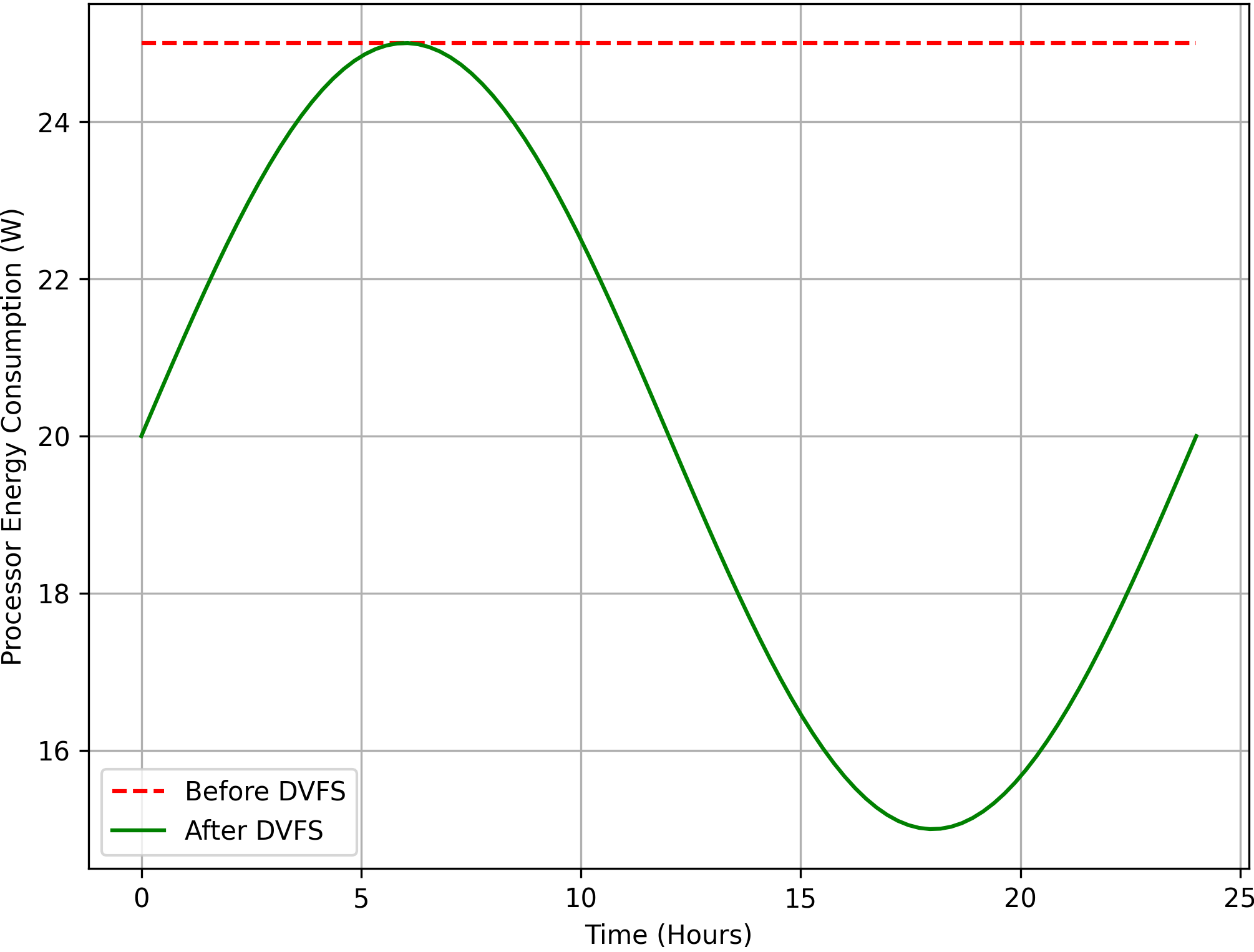}
    \caption{\textbf{Impact of DVFS on processor energy consumption.} Shows a 20\% reduction in energy consumption over time without affecting system responsiveness.}
    \label{fig:dvfs-impact}
\end{figure}

\begin{itemize}
    \item \textbf{Overall Vehicle Energy Consumption}: Through guided incremental optimizations, we achieved a 12\% reduction, decreasing consumption from 50 W to 44 W. This demonstrates the framework's ability to ensure local optimizations contribute to system-wide gains.
    \item \textbf{Processor Energy Consumption}: By applying Dynamic Voltage and Frequency Scaling (DVFS), we reduced the processor's energy consumption by 20\%, as depicted in Figure~\ref{fig:dvfs-impact}. These adjustments were validated at each step to ensure they positively impacted the global system.
    \item \textbf{Communication Module Energy Consumption}: Optimizations in data transmission led to a 25\% reduction in energy usage. Each change was carefully evaluated to confirm its contribution to overall efficiency gains.
    \item \textbf{System Responsiveness}: Despite the energy optimizations, system responsiveness was maintained, with average decision-making latency remaining unchanged. This balance underscores the framework's holistic approach, ensuring that energy efficiency does not come at the expense of performance.
    \item \textbf{Measurement Precision}: The precision parameter allowed us to refine measurement accuracy, reducing error margins to below 5\%. This ensured that our optimization efforts were guided by reliable data.
\end{itemize}

Overall, the GMB-ECC framework demonstrated its capability to guide and validate the optimization process, ensuring that local changes translated into significant global improvements. This approach not only enhances energy efficiency but also supports sustainable and cost-effective operations across complex systems.

\begin{figure}[t]
    \centering
    \includegraphics[width=\linewidth]{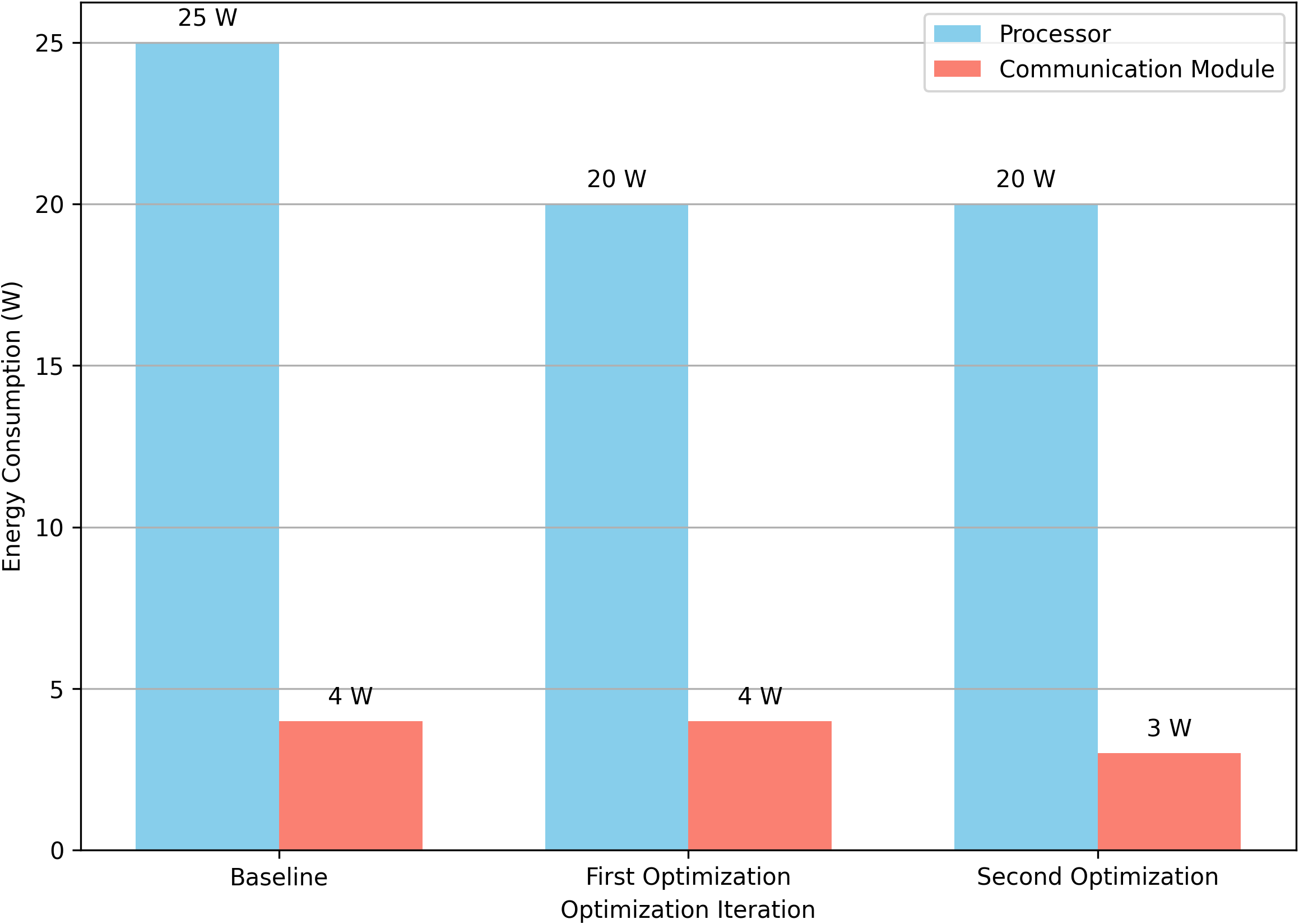}
    \caption{\textbf{Energy consumption reduction across optimization iterations.} Reduction in energy consumption of the processor and communication module across optimization iterations.}
    \label{fig:energy-consumption}
\end{figure}\clearpage{}%
\clearpage{}%
\section{Related Work}\label{sec:related_work}
 It is widely acknowledged that the demand for cloud resources is rapidly increasing, and the resulting increase in energy consumption has become a pressing challenge \cite{Cloudzero.2025}. Policymakers have recognized this issue and are implementing measures to address rising energy usage in cloud computing. For instance, Germany's Energy Efficiency Act sets power usage effectiveness (PUE) limits for data centers \cite{enefg2023}. Meanwhile, researchers are actively exploring various solutions to reduce energy consumption and enhance energy efficiency in cloud data centers \cite{Katal2023, Alharbi2019, Panwar.2022}. However, energy consumption in the edge-cloud continuum has received considerably less attention in the research community. Optimizing applications globally across the edge remains a significant challenge \cite{Ahvar.2019}. Existing work often focus on specific applications, limiting their broader applicability, such as smart geriatric homes \cite{Swain.2022}, smart connected health \cite{Materwala.2021}, or human digital twins \cite{Zhang.2024b}. Other contributions target entire sectors, like vehicular networks \cite{Alahmadi.2020, Gu.2021} or industrial manufacturing \cite{Aouedi.2024, BinMofidul.2022}. Additionally, some research addresses individual components, such as end devices \cite{Li.2024f, Tayade.2017}. Most contributions propose novel, optimized methods for solving specific problems, such as task offloading \cite{Wu.2021, Li.2024f, ElHaber.2019, Aujla.2017, Chen.2022b}, or scheduling \cite{Kaur.2020, Alatoun.2022, Li.2023c}. However, few studies take a holistic view of energy efficiency across the entire system. For example, while \cite{Patel.2024} adopts a broad perspective, it does not provide detailed insights into the underlying computational mechanisms. 

\textbf{End-to-End Energy Measurement and Optimization in Edge-Cloud Continuum.}  Li et al. \cite{Li.2018} provide an end-to-end analysis for video streaming, using static models and assumptions. While insightful, this approach lacks adaptability across diverse applications. GMB-ECC extends this approach, by offers a framework that spans the entire continuum, integrating versatile practices with a precision parameter for adaptable measurement granularity. This adaptability supports practitioners in selecting appropriate methods, enhancing energy efficiency without extensive infrastructure modifications.

\textbf{Energy Management in Data Centers } Aujla and Kumar \cite{Aujla.2017} developed an energy management scheme for data centers in edge cloud environment, considering energy consumption as well as dynamic energy provisioning from multiple energy sources. Energy consumption beyond the data center is not considered.

The proposed multiple optimization approach includes workload classification and scheduling for optimal computing resource and network utilization. Contrary to GMB-ECC, their objective was not to ensure comparability of energy efficiency across all optimization approaches.

\textbf{Energy Consumption Modeling and Analysis.} Ahvar et al. \cite{Ahvar.2019} proposed a detailed model differentiating static and dynamic components. They aimed to develop a generic and accurate model. Yet its complexity results in maintenance challenges. GMB-ECC overcomes the differentiation by focusing on seamless integration and practical adaptability, avoiding static modeling pitfalls. This allows effective energy management aligned with the dynamic nature of modern environments, enhancing operational efficiency and scalability.

\textbf{Cost-Effectiveness Measurement Method.} Li et al. \cite{Li.2020} investigate a computational offloading and resource allocation problem in a vehicular edge cloud setting. The optimization targets are time conservation and energy consumption. They propose a value density function that relates energy conservation to the number of resources allocated. A formula for calculating an energy equilibrium point is presented. They use a \textit{beneficial decision vector} to guide through a discrete set of action options for optimal resource allocation. This is an interesting approach, however, the proposed measurement approach is ultimately applicable only to a specific problem.

\textbf{Cost-Benefit Analysis of Iterative Optimization Methods. } For NP-hard problems like service placement and resource allocation, there is no known method to calculate an optimal solution using direct approaches. Therefore, iterative methods are often employed to find approximate solutions. While more iterations may lead to better solutions, this results in higher computational effort, which in turn means more time and energy consumption. Xiang et al. \cite{Xiang.2022} depict, that the relation between optimal energy consumption and optimal performance is not linear, but you need to find a good stop criterion for an optimization algorithm. It may not find the exact minimum of the corresponding cost function, but the solution will be within a certain threshold distance from the optimum.

\textbf{Energy Consumption Estimation Methods.} Qu et al. \cite{Qu.2020} review estimation methods based on CPU utilization, assuming stable environments. GMB-ECC extends these ideas by incorporating less directly measurable components and introducing a precision parameter, allowing for nuanced assessments that accommodate the complexity and variability of edge-cloud environments.

\textbf{Real-Time Energy Data Acquisition and Prediction.} Recent advancements emphasize precise measurement techniques. Ismail and Materwala \cite{Ismail.2024} and Bin Mofidul et al. \cite{BinMofidul.2022} focus on data acquisition and anomaly detection yet lack integration across the continuum. GMB-ECC extends these efforts by prioritizing relevant, cost-effective measurements, allowing iterative model refinement. This ensures continuous optimization and alignment with dynamic system requirements.
\clearpage{}%
\clearpage{}%
\section{Limitations and Future Work}\label{sec:limitations_and_future_work}

\textbf{External Validity.} Our framework has shown promising results in the applicability to an autonomous intra-logistic use case using synthetic data. This data helped us demonstrate the GMB-ECC framework's ability to identify inefficiencies and guide optimization. However, using synthetic data comes with limitations regarding external validity. It may not accurately reflect real-world complexities such as unpredictable workload changes, hardware inconsistencies, and environmental factors. These challenges can affect how well our results apply to operational systems. To fully evaluate the framework's applicability and generalizability, we need further validation with real-world data. Testing GMB-ECC across different use cases and industrial scenarios will help assess its effectiveness in various environments. This will increase confidence in its practical utility and establish its reliability in real-world applications.

\textbf{Internal Validity.} Our framework's internal validity has been maintained throughout the demonstration with our running example. We've ensured that the algorithms and methodologies used are consistent, reliable, and free from biases. However, as we extend the framework's application to additional use cases, it's required to continuously monitor and maintain this internal consistency. This will ensure that the results remain valid and actionable across different scenarios. Ensuring ongoing consistency in the framework's application will be essential to maintain internal validity and produce correct results across various scenarios.

\textbf{Construct Validity.} A notable improvement in our framework is the introduction of a precision parameter, which has been defined to allow the model to self-describe its construct validity. This parameter helps in adjusting the complexity and granularity of the measurements taken, ensuring they align with the heterogeneous nature of the edge-cloud continuum.
By enabling the model to control the level of detail in its representation, the precision parameter ensures that the energy efficiency metrics are both relevant and accurate. This adaptability allows the framework to adjust to varying resource conditions and operational contexts while maintaining a clear connection between the measured data and the underlying constructs of energy efficiency. Standardized metrics for evaluating sustainability further support this construct validity, providing a foundation for benchmarking improvements across different edge-cloud environments.

\textbf{Implications and Future Work.} The use of synthetic data highlights the preliminary nature of our evaluation and underscores the necessity for cautious interpretation of the results. Further validation is needed to confirm the framework's effectiveness in operational settings. Moving forward, we plan to:
\begin{itemize}
    \item Pilot Deployments: Implement the GMB-ECC framework in collaboration with industry partners to gather empirical data from operational edge-cloud environments.
    \item Data Collection and Analysis: Monitor energy consumption across various components under real workloads to refine our energy efficiency models.
    \item Model Refinement: Adjust the precision parameter based on actual measurement challenges encountered in practice, enhancing the accuracy of our assessments.
    \item Scalability Testing: Assess the framework's performance in larger, more complex systems to ensure it remains effective at scale.
\end{itemize}

By conducting these steps, we aim to improve the external validity and construct validity of the GMB-ECC framework. Real-world validation will strengthen confidence in its applicability and effectiveness, allowing us to generalize the findings to a broader range of scenarios.
\clearpage{}%
\clearpage{}%
\section{Lessons Learned}\label{sec:lessons_learned}

\textbf{Don’t Reinvent the Wheel.} Initially, we sought to develop an understanding of energy efficiency in edge-cloud systems by exploring novel methodologies, such as ML modelling, for analyzing hardware components. This approach, while innovative, introduced unnecessary complexity and heightened risk, leading to confusion among stakeholders. Upon reflection, we realized that existing energy efficiency curves and consumption behaviors were already well-defined in the specifications of these components. By aligning our efforts with these established insights, we could achieve a methodology including well established practices and understanding across various stakeholders. This experience reinforced the principle of avoiding unnecessary reinvention, illustrating that leveraging existing knowledge can address energy efficiency challenges if coordinated correctly.

\textbf{Partial Observability.} In our pursuit of a holistic view of the edge-cloud system, we initially assumed naively to capture and measure all elements simultaneously. However, the vast and distributed nature of these systems proved that full observability was unrealistic. This realization prompted us to adopt an incremental approach, focusing on approximating and measuring specific parts of the system over time. By managing our observations, we could gradually build an understanding of the overall system behavior, at lower implementations cost allowing gradual roll out. This experience underscored the importance of recognizing the limitations of our observational capabilities and adapting our strategies accordingly.
\clearpage{}%
\label{sec:lessons_learned}
\clearpage{}%
\section{Conclusion}\label{sec:conclusion}

We introduced GMB-ECC, an extendable framework for measuring and benchmarking energy consumption across heterogeneous edge-cloud environments. By utilizing energy metrics and incorporating the novel precision parameter, our framework dynamically adjusts measurement granularity. This approach provides accurate energy assessments with minimal computational overhead.

Our experimental evaluation in an autonomous intra-logistic use case demonstrated that GMB-ECC guided us achieve a significant 12\% reduction in energy consumption without compromising system performance by evaluating the global use case for targeted optimization. Compared to traditional measurement approaches, GMB-ECC offers use-case specific precision and adaptability, making it a valuable tool for optimizing energy efficiency in complex computing environments.

Future work includes extending GMB-ECC's evaluation to diverse industrial scenarios to assess its generalizability. We also plan to refine the energy models for greater granularity and integrate advanced optimization algorithms to enhance scalability and enable real-time adjustments.
\clearpage{}%
\label{sec:conclusion}

\section*{Acknowledgment}
The authors thank Ralf Nagel, Jan Peter Meyer, Antje Rätzer-Scheibe, Uwe Cohen, Dominik Loroch and Heinrich Pettenpohl for the insightful input throughout the development of this work. We are grateful to T-Systems International GmbH for their funding in the scope of the EU project IPCEI Next Generation Cloud Infrastructure and Services (IPCEI-CIS), which made this research possible. 

No AI tools were used in the creation of this manuscript; all content was developed and refined by the authors.

\bibliographystyle{amsplain}
\bibliography{literature}

\endgroup
\end{document}